# Basic considerations for magnetization dynamics in the combined presence of spin-transfer torques and thermal fluctuations

Neil Smith
San Jose Research Center
Hitachi Global Storage Technologies
San Jose, CA 95123

## Abstract

This article reviews basic theoretical features of magnetization dynamics of a single domain magnetic film in the presence of spin-transfer torques, with and without thermal fluctuations taken into account. Rather than showing results of detailed numerical calculations, the discussion here is restricted to basic analytical results and conclusions which can mostly be derived from simply the *form* of the equations of motion, as well as elementary considerations based on classical stability analysis and the fluctuation-dissipation theorem. The presents work describes how interesting features of spin-transfer may be viewed as arising from *non-equilibrium* thermodynamics that are a direct consequence of the *nonreciprocal* nature of spin-transfer torques. The present article discusses fairly general results for spin-torque induced instability without thermal fluctuations, as well as the case of thermally activated magnetization reversal in uniaxial devices in the combined presence of external fields, thermal fluctuations, and spin-transfer torques. The results will be discussed and briefly compared and contrasted with that of prior work.

## I. INTRODUCTION

Since the initial theoretical predictions[1,2] and first experimental observations,[3-6] the phenomenon of spin-transfer via dc spin-polarized conduction currents has gained wide practical interest, perhaps primarily for application as a switching mechanism for magnetic memory (MRAM) elements. Since that time, the have been several papers describing the basic device physics and magnetization dynamics associated with this phenomenon through combined numerical and/or direct analytical modeling[7-10]. This article is a contribution by this author to the latter category, but which takes a somewhat different approach from most prior efforts. In particular, the conclusions drawn herein come in large part from the examination of the *form* (rather than the detailed solution) of the same or similar equations of motion, combined with application of elementary stability analysis and basic principles of fluctuation-dissipation theory.[11,12] In regards to the particular issue of spin-transfer torques promoting magnetization reversal in uniaxial devices, the present results describe the effect as a fundamentally *nonequlibrium* (thermo)*dynamics* in which a *nonreciprocal* spin-system can systematically absorb energy from the *dc* spin-polarized conduction currents, thereby aiding if not supplanting the thermal contribution required to overcome the energy barrier for reversal. This description, equivalent to a *dynamical* modification by spin-transfer of the energy barriers for reversal, is quite consistent with recent experiments over a range of magnetic fields, bias currents and temperatures.[13]

However, this description somewhat contrasts prior works[8,10,14] which interpret or model the spin-torque influence on the reversal mechanism as modifying the "effective" spin-temperatures and damping parameters in an otherwise conventional *equilibrium* thermodynamics of the spin system. These differences will be commented on further below. The present analysis regarding combined spin-torque and thermally activated reversal bears similarity in starting point and final result with that of Li and Zhang.[11] However, the details of the derivations are quite different, and as discussed below, this author cannot reconcile the latter derivation[11] with these final results.

## II. BASIC EQUATIONS OF MOTION

The physical basis for spin-transfer torques in a multilayer with two magnetic layers carrying a *dc* CPP conduction current was described previously by Slonczewski.[1] For a conventional spin-valve structure with a reference layer of *fixed* magnetization orientation $\hat{\boldsymbol{m}}_{\mathrm{ref}}$ (e.g., via exchange pinning to an antiferromagnet), the spin transfer torque contribution $\Gamma_{\mathrm{st}}$ to $d\hat{\boldsymbol{m}}_{\mathrm{fre}}/dt$ for the motion of the (unit) magnetization $\hat{\boldsymbol{m}}_{\mathrm{fre}}$ of the remaining "free" layer may be expressed as[1]

$$d\hat{\boldsymbol{m}}_{\text{fre}}/dt \leftrightarrow \Gamma_{\text{st}} \equiv \gamma H_{\text{st}}\,\hat{\boldsymbol{m}}_{\text{fre}} \times (\hat{\boldsymbol{m}}_{\text{ref}} \times \hat{\boldsymbol{m}}_{\text{fre}})$$

$$H_{\text{st}} \equiv (3.3\times 10^{-12}\ \text{Oe-emu/mA})\ \beta(\hat{\boldsymbol{m}}_{\text{fre}} \cdot \hat{\boldsymbol{m}}_{\text{ref}})\ P I_b/(M_s V_{\text{fre}}) \tag{1}$$

$$\beta(x) \equiv 8\sqrt{P}\,[(1+P)^2(3+x) - 16P^{3/2}]^{-1}$$

In Eqs. (1), $P$ is the spin polarization of conduction electrons, $I_b$ is the *dc* bias current, and $V_{\text{fre}}$ is the volume of the free layer. *Positive* $I_b$ refers to current flow where conduction *electrons* flow *from* the reference later, *towards* the free layer. The function $\beta(\hat{\boldsymbol{m}}_{\text{fre}} \cdot \hat{\boldsymbol{m}}_{\text{ref}})$ arises from spin-dependent electron transmission/reflection at the FM/NM interfaces between the NM spacer and free and reference layers.[1] The magnitude of the spin-transfer torques is conveniently described in terms of the scalar "spin-torque-field", $H_{\text{st}}(I_b, \hat{\boldsymbol{m}}_{\text{fre}})$ defined in Eq. (1).

The equations of motion, including spin-transfer torque $\Gamma_{\text{st}}$, are readily obtained by adding $\Gamma_{\text{st}}$ from Eq. (1), to the right hand side of the standard Gilbert equations of motion:

$$d\hat{\boldsymbol{m}}_{\text{fre}}/dt = -(\gamma/M_s V_{\text{fre}})(\partial E/\partial\hat{\boldsymbol{m}}_{\text{fre}} \times \hat{\boldsymbol{m}}_{\text{fre}}) + \alpha\,\hat{\boldsymbol{m}}_{\text{fre}} \times d\hat{\boldsymbol{m}}_{\text{fre}}/dt\ + \Gamma_{\text{st}}$$

$$\Rightarrow d\hat{\boldsymbol{m}}_{\text{fre}}/dt = \gamma\boldsymbol{H}_{\text{eff}} \times \hat{\boldsymbol{m}}_{\text{fre}} + \alpha\,\hat{\boldsymbol{m}}_{\text{fre}} \times d\hat{\boldsymbol{m}}_{\text{fre}}/dt \tag{2}$$

$$\boldsymbol{H}_{\text{eff}} = -(1/M_s V_{\text{fre}})\,\partial E/\partial\hat{\boldsymbol{m}}_{\text{fre}} + H_{\text{st}}\,\hat{\boldsymbol{m}}_{\text{fre}} \times \hat{\boldsymbol{m}}_{\text{ref}}$$

where $E$ is the thermodynamic free-energy of the (single-domain) free layer, and $\alpha$ is the Gilbert (intrinsic) damping parameter. Inclusion of a nonzero spin-torque $\Gamma_{\text{st}}$ simply adds an additional term to the "effective field" $\boldsymbol{H}_{\text{eff}}$, as shown in Eq. (2). For simplicity, the free layer magnetization orientation $\hat{\boldsymbol{m}}_{\text{fre}}(t)$ will henceforth be denoted simply as $\hat{\boldsymbol{m}}(t)$ as clarity permits.

To facilitate derivations herein, $E(\hat{\boldsymbol{m}})$ will (as is most often the case) be taken to be a *quadratic* functional of the cartesian components of $\hat{\boldsymbol{m}}$. It follows from Eq. (2) that

$$\boldsymbol{H}_{\text{eff}} = \boldsymbol{H}_a - \overleftrightarrow{\boldsymbol{H}} \cdot \hat{\boldsymbol{m}}$$

$$(\overleftrightarrow{\boldsymbol{H}})_{ij} \equiv \frac{1}{M_s V_{\text{fre}}}\frac{\partial^2 E}{\partial m_i \partial m_j} - H_{\text{st}}\sum_k \varepsilon_{ijk}(\hat{\boldsymbol{m}}_{\text{ref}})_k \quad (i, j, k = x, y, \text{and/or } z) \tag{3}$$

where $\boldsymbol{H}_a$ is a (quasi)-static external applied field, $\varepsilon_{ijk}$ is the well known *antisymmetric* Levi-Civita tensor of tensor calculus, and $\overleftrightarrow{\boldsymbol{H}}$ is a $3\times 3$ cartesian "stiffness-field" tensor[12] discussed further below. Applying the $\hat{\boldsymbol{m}}\times$ vector operation on Eqs. (2), they may be re-expressed

$$\hat{\boldsymbol{m}}(t) \times \frac{d\hat{\boldsymbol{m}}}{dt} = -\alpha\frac{d\hat{\boldsymbol{m}}}{dt}\ + \{\gamma\hat{\boldsymbol{m}} \times (\boldsymbol{H}_{\text{eff}} \times \hat{\boldsymbol{m}}) \equiv \gamma[\boldsymbol{H}_{\text{eff}} - (\boldsymbol{H}_{\text{eff}} \cdot \hat{\boldsymbol{m}})\hat{\boldsymbol{m}}(t)]\} \tag{4}$$

Substituting Eq. (4) into the rightmost damping term of the Gilbert Eqs. (2) would convert them to their commonly used Landau-Lifshitz-Gilbert form.

Equations (4) are now transformed to a rotated (primed) coordinate system $x'y'z'$ in which the magnetization $\hat{\boldsymbol{m}}_0 \equiv \hat{\boldsymbol{m}}(t = t_0)$ lies along the $+\hat{\boldsymbol{z}}'$ axis, i.e. $\hat{\boldsymbol{m}}'(t_0) = \hat{\boldsymbol{z}}'$. In circumstances and/or for times $t \geq t_0$ where $\hat{\boldsymbol{m}}(t)$ does not rotate "far" from $\hat{\boldsymbol{m}}_0$, Eqs. (4) may *linearized* by expressing the 3-*D* vector $\hat{\boldsymbol{m}}(t \geq t_0)$ to *first* order in the components of the 2-*D* vector $\boldsymbol{m}'(t) \equiv \begin{pmatrix} m'_x(t) \\ m'_y(t) \end{pmatrix}$. With reference to Fig. 1, this transformation is explicitly given by[12]

$$\hat{\boldsymbol{m}} \equiv \begin{pmatrix} m_x \\ m_y \\ m_z \end{pmatrix} = \hat{\boldsymbol{m}}_0 + \overleftrightarrow{\boldsymbol{R}} \cdot \boldsymbol{m}', \;\; \hat{\boldsymbol{m}}_0 \equiv \begin{pmatrix} \sin\theta_0 \\ \cos\theta_0 \sin\phi_0 \\ \cos\theta_0 \cos\phi_0 \end{pmatrix}, \;\; \overleftrightarrow{\boldsymbol{R}} \equiv \begin{pmatrix} \cos\theta_0 & 0 \\ -\sin\theta_0 \sin\phi_0 & \cos\phi_0 \\ -\sin\theta_0 \cos\phi_0 & -\sin\phi_0 \end{pmatrix} \tag{5}$$

Nonzero variations in the "small" $\boldsymbol{m}'(t)$ are dynamically driven by an analogously "small", perturbation field $\boldsymbol{h}'(t)$ (now included separate from $\boldsymbol{H}'_a$) which will later be interpreted as a random *thermal* field $\boldsymbol{h}'_{\text{th}}(t)$. The transformed, *linearized* Eqs. 4 can be re-expressed as

$$\hat{\boldsymbol{z}}' \times \frac{d\boldsymbol{m}'}{dt} + \alpha \frac{d\boldsymbol{m}'}{dt} = \gamma[\boldsymbol{h}'(t) + \boldsymbol{H}'_{\text{eff}} - (\boldsymbol{H}_{\text{eff}} \cdot \hat{\boldsymbol{m}})_0 \, \boldsymbol{m}'(t)] \tag{6}$$

In Eq. (6), the scalar quantity $(\boldsymbol{H}_{\text{eff}} \cdot \hat{\boldsymbol{m}})_0$ need be expressed only to *zeroth* order in $\boldsymbol{m}'$, and makes use of the rotational invariance: $\boldsymbol{H}_{\text{eff}} \cdot \hat{\boldsymbol{m}} = \boldsymbol{H}'_{\text{eff}} \cdot \hat{\boldsymbol{z}}'$, where $\boldsymbol{H}'_{\text{eff}} \equiv \overleftrightarrow{\boldsymbol{R}}^{\mathsf{T}} \cdot \boldsymbol{H}_{\text{eff}}$, with $\overleftrightarrow{\boldsymbol{R}}^{\mathsf{T}}$ the transpose of rotation matrix $\overleftrightarrow{\boldsymbol{R}}$ (Eq. 5). Combining Eqs.(3-6) yields the result:

$$\frac{d}{dt}\left[\frac{1}{\gamma}\begin{pmatrix} \alpha & -1 \\ 1 & \alpha \end{pmatrix} \cdot \begin{pmatrix} m'_x \\ m'_y \end{pmatrix}\right] + \begin{pmatrix} H'_{xx} & H'_{xy} \\ H'_{yx} & H'_{yy} \end{pmatrix} \cdot \begin{pmatrix} m'_x \\ m'_y \end{pmatrix} = \begin{pmatrix} h'_x(t) \\ h'_y(t) \end{pmatrix} + \begin{pmatrix} H'_{ax} \\ H'_{ay} \end{pmatrix}$$

$$\boldsymbol{H}'_a \equiv \overleftrightarrow{\boldsymbol{R}}^{\mathsf{T}} \cdot (\boldsymbol{H}_a - \overleftrightarrow{\boldsymbol{H}} \cdot \hat{\boldsymbol{m}}_0) \tag{7}$$

$$\overleftrightarrow{\boldsymbol{H}}' \equiv \overleftrightarrow{\boldsymbol{R}}^{\mathsf{T}} \cdot \overleftrightarrow{\boldsymbol{H}} \cdot \overleftrightarrow{\boldsymbol{R}} + \begin{pmatrix} (\boldsymbol{H}_a - \overleftrightarrow{\boldsymbol{H}} \cdot \hat{\boldsymbol{m}}_0) \cdot \hat{\boldsymbol{m}}_0 & 0 \\ 0 & (\boldsymbol{H}_a - \overleftrightarrow{\boldsymbol{H}} \cdot \hat{\boldsymbol{m}}_0) \cdot \hat{\boldsymbol{m}}_0 \end{pmatrix}$$

where $\overleftrightarrow{\boldsymbol{H}}$ is given in Eqs. (3). For the results below, the $\boldsymbol{m}'$ dependence of the spin-torque field $H_{\text{st}}$ (see Eq. (1)) will be ignored for simplicity. For the special but important collinear case where $\hat{\boldsymbol{m}}_{\text{ref}} \cdot \hat{\boldsymbol{m}}_0 = \pm 1$, the dependence is only *second* order in $m'_x, m'_y$.

## III. FLUCTUATION-DISSIPATION CONSIDERATIONS

Consider firstly the important case where $\hat{\boldsymbol{m}}_0$ in Eqs. (7) is a *stable equilibrium*, such that $\hat{\boldsymbol{m}} \to \hat{\boldsymbol{m}}_0 \Leftrightarrow H'_{ax} = H'_{ay} = 0$. In this circumstance Eqs (7) have the matrix form:

$$(\overleftrightarrow{\boldsymbol{D}}' + \overleftrightarrow{\boldsymbol{G}}') \cdot (d\boldsymbol{m}'/dt) + \overleftrightarrow{\boldsymbol{H}}' \cdot \boldsymbol{m}'(t) = \boldsymbol{h}'(t)$$

$$\overleftrightarrow{\boldsymbol{D}}' \equiv \frac{1}{\gamma}\begin{pmatrix} \alpha & 0 \\ 0 & \alpha \end{pmatrix}, \ \overleftrightarrow{\boldsymbol{G}}' = \frac{1}{\gamma}\begin{pmatrix} 0 & -1 \\ 1 & 0 \end{pmatrix} \qquad (8)$$

with $\overleftrightarrow{\boldsymbol{H}}'$ given by Eqs. (3,7). *In the absence of spin-transfer torques* (i.e., $H_{\text{st}} = 0$), the stiffness tensor $\overleftrightarrow{\boldsymbol{H}}' \propto \partial^2 E / \partial \boldsymbol{m}' \partial \boldsymbol{m}'$ *is a symmetric positive definite matrix,* and Eqs. (8) have previously been shown[12] to be of a canonical form describing the equilibrium dynamics a linear system of coupled, damped harmonic oscillators, here $m'_x(t)$ and $m'_y(t)$, which fluctuate about a *thermal equilibrium* $\langle m'_x \rangle = \langle m'_y \rangle = 0$. The fluctuations in $\boldsymbol{m}'(\boldsymbol{t})$ may be interpreted as the dynamical response (as governed by Eqs. (8)) of $\boldsymbol{m}'(\boldsymbol{t})$ to *random* "thermal fields" $\boldsymbol{h}'(\boldsymbol{t}) \to \boldsymbol{h}'_{\text{th}}(t)$ (with $\langle \boldsymbol{h}'_{\text{th}} \rangle = 0$). As described in the Appendix 1, the fluctuation-dissipation theorem shows that the statistical properties of the $\boldsymbol{h}'_{\text{th}}(t)$ are determined *solely* from the *symmetric* "damping matrix" $\overleftrightarrow{\boldsymbol{D}}'$ of Eqs.(8). In particular, the correlation matrix $\langle \boldsymbol{h}'_{\text{th}}(0) \boldsymbol{h}'_{\text{th}}(\tau) \rangle$ is given by[12]

$$\begin{pmatrix} \langle h'_x(0) h'_x(\tau) \rangle & \langle h'_x(0) h'_y(\tau) \rangle \\ \langle h'_y(0) h'_x(\tau) \rangle & \langle h'_y(0) h'_y(\tau) \rangle \end{pmatrix} = \frac{2k_B T}{M_s V_{\text{fre}}} \overleftrightarrow{\boldsymbol{D}}' \delta(\tau) = \frac{2k_B T}{\gamma M_s V_{\text{fre}}} \begin{pmatrix} \alpha & 0 \\ 0 & \alpha \end{pmatrix} \delta(\tau) \qquad (9)$$

By contrast, if $H_{\text{st}} \neq 0$, the *non-conservative* spin-torque contribution $H_{\text{st}} \hat{\boldsymbol{m}}_{\text{ref}} \times \hat{\boldsymbol{m}}_{\text{fre}}$ to $\boldsymbol{H}_{\text{eff}}$ (Eq. (2)) is *not* derivable as the the gradient $\partial E / \partial \boldsymbol{m}$ of a free energy. Similarly, both $\overleftrightarrow{\boldsymbol{H}}$ and $\overleftrightarrow{\boldsymbol{H}}'$ tensors (see Eqs. (3,7)) becomes *nonreciprocal* (e.g., $\overleftrightarrow{H}'_{xy} \neq \overleftrightarrow{H}'_{yx}$) when $H_{st} > 0$, and are also *not* of the form $\partial^2 E / \partial \boldsymbol{m} \partial \boldsymbol{m}$. Hence, in the presence of spin-transfer torques from polarized conduction currents, the magnetization/spins are no longer in strict *thermal* equilibrium with the lattice "thermal bath". As is discussed below, the *nonreciprocal* nature of $\overleftrightarrow{\boldsymbol{H}}'$ implies the possibility of systematic (nonrandom) energy transfer between the magnetization/spins and the *dc* spin-polarized conduction current, a phenomenon also discussed previously.[2,15]

However, within the context of the Gilbert-Slonczewski description of spin-torque effects as in Eqs. (3-7), the spin-torques do not fundamentally alter the physical mechanisms (e.g., spin-orbit coupling) of energy transfer (or "friction") between magnetization/spins and lattice that are phenomenologically represented via the damping parameter $\alpha$ and the random thermal fields $\boldsymbol{h}'_{\mathrm{th}}(t)$ described stochastically by Eqs. (9). Alternatively stated, with reference to Eq. (8), *the addition of spin-transfer torques* $\mathrm{G}_{\mathrm{st}}$ *modify the stiffness matrix* $\overleftrightarrow{\boldsymbol{H}}'$, *but do not alter the damping matrix* $\overleftrightarrow{\boldsymbol{D}}'$. Such considerations, based on the *form* of the Gilbert-Slonczewski (linearized) equations of motion (Eqs. (7)), are in this author's view, sufficient to show that prior interpretations[8,10,14] for the equivalence of spin-transfer torques with modifications to the spin-temperatures and damping parameter are *incorrect* at a basic *physical* level, even if there may be some mathematical similarities. This point will be discussed further below.

## IV. DYNAMIC STABILITY ANALYSIS WITHOUT THERMAL FLUCTUATIONS

Consider again the case where $\hat{\boldsymbol{m}}_0$ in Eqs. (7) is an *equilibrium* point, such that $\hat{\boldsymbol{m}} \to \hat{\boldsymbol{m}}_0 \Leftrightarrow H'_{ax} = H'_{ay} = 0$. For a perturbation field of the form: $\boldsymbol{h}'(t) \to \boldsymbol{h}'(\omega)e^{-i\omega t}$ with possibly *complex* $\omega$, solutions of the form $\boldsymbol{m}'(t) = \boldsymbol{m}'(\omega)e^{-i\omega t}$ can (if they exist) be found by simple substitution of these expressions into Eqs. (7). The solutions are readily found to be

$$\boldsymbol{m}'(\omega) = \overleftrightarrow{\chi}(\omega)\cdot\boldsymbol{h}'(\omega) \Leftrightarrow \boldsymbol{h}'(\omega) = \overleftrightarrow{\Lambda}(\omega)\cdot\boldsymbol{m}'(\omega)$$
$$\overleftrightarrow{\Lambda}(\omega) = \begin{pmatrix} H'_{xx} - i\omega\alpha/\gamma & H'_{xy} + i\omega/\gamma \\ H'_{yx} - i\omega/\gamma & H'_{yy} - i\omega\alpha/\gamma \end{pmatrix} \tag{10}$$

where $\overleftrightarrow{\chi}(\omega) \equiv \overleftrightarrow{\Lambda}^{-1}(\omega)$ and $\overleftrightarrow{\Lambda}(\omega)$ are (small-signal) susceptibility and inverse susceptibility matrices, respectively.

To examine the *dynamic stability* of the solutions of Eqs. (7,10), one consider the "natural" modes of the system, which by definition satisfy $\det\overleftrightarrow{\Lambda}(\omega) = 0$. The latter condition permits "spontaneous" solutions of *finite* $\boldsymbol{m}'(\omega)$ even as $\boldsymbol{h}'(\omega) \to 0$. The condition $\det\overleftrightarrow{\Lambda}(\omega) = 0$ readily yields the following characteristic equation and solutions for the "natural" frequencies $\omega_n$:

$$(1+\alpha^2)\,\omega_n^2 + i\gamma[\alpha(H'_{xx}+H'_{yy}) - (H'_{xy}-H'_{yx})]\omega_n - \gamma^2 \det\overleftrightarrow{\boldsymbol{H}}' = 0$$
$$\omega_n = \frac{-i\gamma}{2}[\alpha(H'_{xx}+H'_{yy}) - (H'_{xy}-H'_{yx})] \pm \frac{\gamma}{2}\sqrt{4\det\overleftrightarrow{\boldsymbol{H}}' - [\alpha(H'_{xx}+H'_{yy}) - (H'_{xy}-H'_{yx})]^2} \tag{11}$$

approximating $(1+\alpha^2) \cong 1$ in the last result. Since the natural modes can become excited at the *slightest* (thermal) perturbation of the system, the dynamic stability criterion is that any natural mode $\boldsymbol{m}'_n(t) \propto e^{-i\omega_n t}$ *must decay with time, which requires that* $\mathrm{Im}\,\omega_n < 0$. Hence, an equilibrium point $\hat{\boldsymbol{m}}_0$ will be *stable* when *both* of the following conditions hold:

$$\det \overleftrightarrow{\boldsymbol{H}}' \equiv H'_{xx}H'_{yy} - H'_{xy}H'_{yx} \; > \; 0 \tag{12}$$

$$\alpha(H'_{xx} + H'_{yy}) + (H'_{yx} - H'_{xy}) \; > \; 0 \tag{13}$$

The stability criteria of Eqs. (12,13) both broadly generalize *and simplify* the results of Grollier et. al[9], which were restricted only to cases $\hat{\boldsymbol{m}}_0 \cdot \hat{\boldsymbol{m}}_{\mathrm{ref}} = \pm 1$, and employed additional *unnecessary* assumptions. This latter flaw in that analysis[9] appears to stem from a premature dropping of the term $\alpha^2 (H'_{xx} + H'_{yy})^2$ from the expression $[\alpha(H'_{xx}+H'_{yy}) - (H'_{xy}-H'_{yx})]^2$ inside the discriminent ($\sqrt{\;}$ term) in Eq. (11) because it is of order $\alpha^2$. In fact, both terms inside the above $[\;]^2$ in Eq. (11) are of comparable size in cases of practical interest (e.g., Eq. (13)).

Without spin-torques ($H_{\mathrm{st}} = 0$), the spin/magnetization system described by Eqs. (3,7) is *reciprocal*, i.e., $H'_{xy} = H'_{yx}$ .In *this* case one can always rotate the $x', y'$ axes to "principle-axes" such that $H'_{xy} = H'_{yx} = 0$. Hence, for reciprocal systems, the *sole* stability criterion is Eq. (12), $\det \overleftrightarrow{\boldsymbol{H}}' > 0 \Rightarrow H'_{xx}H'_{yy} > 0$. However, the *nonreciprocal* nature of the spin-torque contribution to $\overleftrightarrow{\boldsymbol{H}}'$ *unconditionally requires* that $H'_{xy} \neq H'_{yx}$ ($H'_{xy} = -H'_{yx} \propto H_{\mathrm{st}}$ for principle-axes), and hence introduces the *additional* stability criterion of Eq. (13). Here, an otherwise stable equilibrium point *becomes* unstable above critical values of $|H_{\mathrm{st}}|$. This corresponds to the spin-torque induced magnetization observed experimentally.[3-6] Nonreciprocal spin-torques permit yet another possibility, that of a *positive* contribution $-H'_{xy}H'_{yx}$ in Eq. (12), such that a sufficiently large $|H_{\mathrm{st}}|$ can *stabilize* an otherwise unstable equilibrium point with $H'_{xx}H'_{yy} < 0$.

The physical consequences of a nonreciprocal stiffness tensor $\overleftrightarrow{\boldsymbol{H}}'$ may be further elucidated by the following thermodynamical considerations. The internal work $\Delta W$ done *by* (and/or internal energy *lost of*) the free-layer in going from an initial magnetization state $\hat{\boldsymbol{m}}(t = t_0 = 0) = \hat{\boldsymbol{m}}_0$, to the magnetization state $\hat{\boldsymbol{m}}_1(t)$ is given by[16]

$$\Delta W / M_s V_{\mathrm{fre}} \equiv -\int_{\hat{\boldsymbol{m}}_0}^{\hat{\boldsymbol{m}}_1} \boldsymbol{H}_{\mathrm{eff}} \cdot d\hat{\boldsymbol{m}} = -\int_0^{t_1} \boldsymbol{H}_{\mathrm{eff}} \cdot (d\hat{\boldsymbol{m}}/dt)\,dt \;\cong -\int_0^{t_1} \boldsymbol{H}'_{\mathrm{eff}} \cdot (d\boldsymbol{m}'/dt)\,dt \tag{14}$$

For *conservative* systems with $\boldsymbol{H}_{\text{eff}} = -\partial E / \partial \hat{\boldsymbol{m}}$, Eq. (14) shows that $\Delta W = E(\hat{\boldsymbol{m}}_1) - E(\hat{\boldsymbol{m}}_0)$ is *independent* of time $t_1$ or the orbital path from from $\hat{\boldsymbol{m}}_0$ to $\hat{\boldsymbol{m}}_1(t)$. For *nonconservative* systems, consider the linear approximation on the right of Eq. (14) along with a hypothetical path resembling a steady state orbital precession of the magnetization of the form (see also Eq. (5)):

$$\hat{\boldsymbol{m}}(t) = \hat{\boldsymbol{m}}_0 + \overleftrightarrow{\boldsymbol{R}} \cdot \boldsymbol{m}'(t) \ , \ \boldsymbol{m}'(t) = \text{Im}\, \boldsymbol{m}_1' e^{i\omega t} \ , \ \boldsymbol{m}_1' = \begin{pmatrix} m_{1x}' \\ m_{1y}' \end{pmatrix} \tag{15}$$

with *real* frequency $\omega \neq 0$. Combining Eqs. (14,15) , along with $\boldsymbol{H}_{\text{eff}}'(t) = \boldsymbol{H}_a' - \overleftrightarrow{\boldsymbol{H}}' \cdot \boldsymbol{m}'(t)$ from Eqs. (3,7), the *time-averaged* work $\overline{dW/dt}$ per cycle ($t_1 = 2\pi/\omega$) is given by

$$\begin{aligned} \overline{dW/dt}\,/M_s V_{\text{fre}} &= -\tfrac{\omega}{2\pi}\int_0^{2\pi/\omega} (\boldsymbol{H}_{\text{eff}}' \cdot d\boldsymbol{m}'/dt)\, dt = \tfrac{\omega}{2\pi}\int_0^{2\pi/\omega} \text{Im}[i\omega \boldsymbol{m}_1' e^{i\omega t}] \cdot \text{Im}[\overleftrightarrow{\boldsymbol{H}}' \cdot \boldsymbol{m}_1' e^{i\omega t}]\, dt \\ &= \tfrac{\omega}{2}\text{Im}[\boldsymbol{m}_1'^* \cdot \overleftrightarrow{\boldsymbol{H}}' \cdot \boldsymbol{m}_1'] = \tfrac{\omega}{2}\text{Im}[m_{1x}' m_{1y}'^*](H_{yx}' - H_{xy}') \end{aligned} \tag{16}$$

Equation (16) shows that *only* a *nonreciprocal* component of the stiffness-field matrix $\overleftrightarrow{\boldsymbol{H}}'$ permits the possibility of a *sustained absorption* of energy (i.e., $\overline{dW/dt} < 0$) by the free-layer's spin/magnetization system by *sole* means of a quasi-periodic motion $\hat{\boldsymbol{m}}(t)$ in the presence of a *stationary* external field $\boldsymbol{H}_a$. In the case of spin-torques, this absorbed energy ultimately comes from the electrical source maintaining the *dc* bias current, and is the physical origin of the spin-torque induced instability described by Eq. (13) (see also Appendix 2). The *polarity* of $\overline{dW/dt}$ varies as $\text{sgn}(H_{st}) \cdot \text{sgn}(\text{Im}[m_{1x}' m_{1y}'^*])$, and thus depends on the *phase and chirality* of the orbital precession described by Eqs. (15). This feature of Eq. (16) is fundamentally distinct from a viscous/frictional Gilbert type damping (the latter representing a thermally equilibrating spin/magnetization-lattice energy transfer process), and cannot be physically or mathematically characterized strictly in terms of "effective" damping parameters invoked previously.[8,10]

## V. DYNAMIC STABILITY ANALYSIS WITH THERMAL FLUCTUATIONS

The stability criteria of Eqs. (12,13) exclude considerations of thermal activation. In particular, if $\hat{\boldsymbol{m}}_1$ in Eq. (14) is an energy maximum "between" minima $\hat{\boldsymbol{m}}_0$ and $\hat{\boldsymbol{m}}_2$, $\Delta W \to W_{\text{act}}$ is the activation energy[11] required to complete the process $\hat{\boldsymbol{m}}_0 \to \hat{\boldsymbol{m}}_1 \to \hat{\boldsymbol{m}}_2$. This energy need be supplied thermally via stochastic energy transfer between the lattice/thermal-bath and the spins/free-layer. The probability $p(t)$ that the transition $\hat{\boldsymbol{m}}_0 \to \hat{\boldsymbol{m}}_2$ takes place with a time $t$ follows an Arrhenius law, $p(t) \cong 1 - e^{-t/\tau}$, $\tau \propto \exp(W_{\text{act}} / k_B T)$.[11,17] However, the presence of spin-torques can significantly reduce the magnitude of $W_{\text{act}}$ ,and hence greatly influence the probability for a thermally activated transition.

For a more specific evaluation, one can consider the following simple, prototypical example. With reference to Fig. 1, this uniaxial case is described by:

$$\boldsymbol{H}_a = -H_a \hat{\boldsymbol{z}},\ \hat{\boldsymbol{m}}_0 = +\hat{\boldsymbol{z}},\ \hat{\boldsymbol{m}}_2 = -\hat{\boldsymbol{z}},\ \hat{\boldsymbol{m}}_{\text{ref}} = \pm\hat{\boldsymbol{z}}$$

$$E / M_s V_{\text{fre}} = \tfrac{1}{2} H_\perp m_y^2 - \tfrac{1}{2} H_k m_z^2 + H_a\, m_z \tag{17}$$

$$H'_{xx} = H_k - H_a\ ,\ H'_{yy} = H_k - H_a + H_\perp\ ,\ H'_{yx} = -H'_{xy} = \pm H_{\text{st}}$$

$H_k$ is an *in-plane* uniaxial anisotropy constant (easy-axis $= \pm\hat{\boldsymbol{z}}$). It will be further specified here that $0 \le H_a < H_k$ so that $\hat{\boldsymbol{m}}_0 = \hat{\boldsymbol{z}}$ is a *thermally stable* equilibrium The out-of-plane anisotropy $H_\perp$ is primarily the shape-demagnetizing field $H_\perp \cong B_s \equiv 4\pi M_s$. In most practical circumstances, $H_\perp \gg |H_a| + |H_k|$, and the motion $\hat{\boldsymbol{m}}(t)$ is primarily in the *x*-z plane, i.e., $m_y(t) \ll m_x(t)$. With reference to Eq. (13,17), it will also be assumed that $|H_{\text{st}}| < \alpha(H_\perp/2 + H_k - H_a)$, so that the state $\hat{\boldsymbol{m}}_0$ is also stable with respect to spin-transfer torques. For the simple collinear, axisymmetric case $\hat{\boldsymbol{m}}_0 = \hat{\boldsymbol{z}} = \pm\hat{\boldsymbol{m}}_{\text{ref}}$ described by Eq. (17), $m_{x,y}(t) \leftarrow m'_{x,y}(t)$, $(\boldsymbol{H}_{\text{st}} \equiv H_{\text{st}} \hat{\boldsymbol{m}} \times \hat{\boldsymbol{m}}_{\text{ref}}) \cdot \hat{\boldsymbol{z}} = 0$, and thus Eqs. (16,20) hold *without* restriction to the linear approximation $m_x^2 + m_y^2 \ll 1$.[18]

As discussed in Appendix 1, thermal fluctuations of the free-layer magnetization $\hat{\boldsymbol{m}}(t)$ roughly resemble a precessional motion (e.g., Eqs. (15)) with resonance frequency $\omega \to \omega_0 \cong \gamma\sqrt{H'_{xx} H'_{yy}}$ , but which remains *phase-coherent* only over *limited* time intervals $\tau_c \cong [\alpha\gamma(H'_{xx} + H'_{yy})/2]^{-1}$ (with $\omega_0 \tau_c \gg 1$ typically). The relationship between $m_x(t)$ and $m_y(t)$ is obtained from the *y*-component of the *nonlinear* Gilbert Eqs. (2,4) and Eq.(17):

$$\frac{dm_y}{dt} + \gamma[\pm H_{\text{st}} \overline{m}_z + \alpha(H_\perp + H_k \overline{m}_z - H_a)] m_y(t) \cong \gamma[H_k \overline{m}_z - H_a] m_x(t) \tag{18}$$

In Eq. (18), $m_z \to \overline{m}_z \approx \overline{(1 - m_x^2)^{1/2}}$ is replaced by a time-averaged value to facilitate an approximate solution of this nonlinear equation. Also, small terms $\alpha H_\perp m_y^2$ and $\alpha H_{\text{st}}$ have been dropped from the [ ] factors on the left and right sides of the equation, respectively. The thermal fields $\boldsymbol{h}'_{\rm th}(t)$ are included implicitly as the driving force for the in-plane motion $m_x(t) \gg m_y(t)$, e.g., the orbit described by Eq. (15). In turn, $m_x(t)$ on the right side of Eq. (18) is treated as the driving source term for the motion $m_y(t)$.

Substituting a solution of the form of Eq. (15) into Eq. (18) then yields the result

$$m_{1y} = \frac{\omega_x(\omega_y - i\omega_0)}{\omega_0^2 + \omega_y^2} m_{1x}$$

$$\omega_x \equiv \gamma[H_k \overline{m}_z - H_a], \;\; \omega_y \equiv \gamma[\pm H_{\rm st}\overline{m}_z + \alpha(H_\perp + H_k\overline{m}_z - H_a)] \quad (19)$$

$$\omega_0 = \gamma\sqrt{(H_k - H_a)(H_\perp + H_k - H_a)}, \;\tau_c = [\alpha\gamma(H_\perp/2 + H_k - H_a)]^{-1}$$

expressing complex amplitude $m_{1y}$ in terms of $m_{1x}$. The result of Eq. (19) implicitly assume that $\omega_0\tau_c \gg 1$ (see Appendix 1), such that one or more quasi-periodic orbits occur over the coherence time $\tau_c$.(That $\omega_0\tau_c \gg 1$ also implies $\omega_0 \gg \omega_y$ if $|H_{\rm st}| < \alpha(H_\perp/2 + H_a + H_k)$. *The path-dependent integrated spin-torque contribution to* $\Delta W$ *is now approximated as* $\overline{dW/dt}\,\tau_c$ (on average). Combining this, Eqs. (14,16,17,19), and replacing $m_{1x} \to \sin\theta_1$, one obtains

$$\begin{aligned}\Delta W / M_s V_{\rm fre} &\cong \tfrac{1}{2} H_k \sin^2\theta_1 - H_a(1-\cos\theta_1) \pm H_{\rm st}\,\omega_x\tau_c \sin^2\theta_1 \\ &\to \tfrac{1}{2} H_k \sin^2\theta_1 - H_a(1-\cos\theta_1) \pm \frac{H_{\rm st}(H_k\overline{m}_z - H_a)}{\alpha(H_\perp/2 + H_k - H_a)}\sin^2\theta_1\end{aligned} \quad (20)$$

The activation energy $W_{\rm act}$ can (in principle) be obtained from Eq. (20) by evaluation at the angle $\theta_{1\max}$ that *maximizes* $\Delta W$. However, this is substantially complicated by the implicit dependence of the time/orbit-averaged $\overline{m}_z$ on $\theta_1(H_a, H_k)$. From limiting cases $\overline{m}_z \to 1$ as $H_a \to H_k \Rightarrow \theta_{1\max} \to 0$, as well as $\overline{m}_z \approx 1/2$ when $H_a \to 0 \Rightarrow \theta_{1\max} \to \pi/2$, one can approximate $H_k\overline{m}_z - H_a \approx \tfrac{1}{2}(H_k - H_a)$. Substituting the latter into Eq. (20) allows a straightforward, analytical maximization of $\Delta W$, which yields the result

$$\begin{aligned} W_{\rm act} &\cong \tfrac{1}{2} H_k^{\rm eff} M_s V_{\rm fre}\, (1 - H_a / H_k^{\rm eff})^2 \;\; (0 \le H_a < H_k) \\ H_k^{\rm eff} &\equiv H_k \pm (H_k - H_a)(H_{\rm st} / H_{\rm st}^{\rm crit}),\; H_{\rm st}^{\rm crit} \equiv \alpha(H_\perp/2 + H_k - H_a)\end{aligned} \quad (21)$$

A heuristic, a posteriori justification for the prior $\overline{m}_z$-approximation is the simple physical interpretation of Eq. (21), which stems from the appearance in Eq. (20) of the critical spin-torque field $H_{\rm st}^{\rm crit}$ as first obtained in Eq. (13). In the most elementary case $H_a = 0$, $W_{\rm act}(H_{\rm st}) = \Delta W(\theta_1 = \pi/2) = \frac{1}{2} M_s V_{\rm fre} H_k^{\rm eff}(H_{\rm st}) = W_{\rm act}(0)[1 \pm (H_{\rm st}/H_{\rm st}^{\rm crit})]$ has a particularly simple form which agrees with prior results.[11] The Boltzmann factor $\exp(W_{\rm act}/k_B T)$ may also be *mathematically* expressed as $\exp(W_{\rm act}(0)/k_B T_{\rm eff})$ with an effective spin-temperature[8,10] $T_{\rm eff}(H_{\rm st}) = T/[1 \pm (H_{\rm st}/H_{\rm st}^{\rm crit})]$ However, the result of Eq. (21) indicates that this simple factorization does *not* generally hold for $H_a > 0$. In all cases, though, $W_{\rm act} \to 0$ as $H_{\rm st} \to \mp H_{\rm st}^{\rm crit}$, and there is no additional required thermal activation energy for free-layer easy-axis reversal: $\hat{\boldsymbol{m}}_0 = \hat{\boldsymbol{z}} \to \hat{\boldsymbol{m}}_2 = -\hat{\boldsymbol{z}}$, once the spin-torque field reaches the critical values $\mp H_{\rm st}^{\rm crit}$. Rather than imply a nonsensical $T_{\rm eff} \to \infty$, the latter merely indicates a *dynamical* reversal process due to energy absorption from spin-torques, as discussed in Sec. IV and Appendix 2.

A related treatment of thermal activation with finite spin-transfer torques was previously described by Li and Zhang.[11] These authors claim a final result that is identical to Eq. (21), in the case $H_a = 0$. However, this author is unable to verify this based on Li and Zhang's explicit description of their own derivation,[11] which this author additionally finds to be faulty. In the present notation, and from Eq. (14), the derivation of the spin-torque contribution $\Delta W_{\rm st}$ to the activation energy , as described step by step by Li and Zhang,[11] is as follows:

$$\Delta W_{\rm st}/M_s V_{\rm fre} \equiv -\int_0^{t_1}[\boldsymbol{H}_{\rm st}\cdot d\hat{\boldsymbol{m}}/dt]dt = -H_{st}\int_0^{t_1}[(\hat{\boldsymbol{m}}\times\hat{\boldsymbol{m}}_{\rm ref})\cdot d\hat{\boldsymbol{m}}/dt]dt \tag{22a}$$

$$= \pm H_{\rm st}\int_0^{t_1}[(\hat{\boldsymbol{m}}\times d\hat{\boldsymbol{m}}/dt)\cdot\hat{\boldsymbol{z}}]dt \to \pm\frac{H_{\rm st}}{\alpha}\int_0^{t_1}[\frac{dm_z}{dt} - \gamma(\boldsymbol{H}_{\rm eff}\times\hat{\boldsymbol{m}})\cdot\hat{\boldsymbol{z}}]dt \tag{22b}$$

$$\to \pm\frac{H_{\rm st}}{\alpha}\Big[(\cos\theta_1 - 1) - \gamma\int_0^{t_1}[\pm H_{\rm st}(m_x^2 + m_y^2) + B_s m_x m_y]dt\Big] \tag{22c}$$

$$\to \pm\frac{H_{\rm st}}{\alpha}\Big[(\cos\theta_1 - 1) + \int_0^{m_{x1}}\frac{(B_s m_y \pm H_{\rm st}m_x)m_x}{\overline{m}_z(B_s m_y \pm H_{\rm st}m_x) + (H_k\overline{m}_z - H_a)m_y}dm_x\Big] \tag{22d}$$

$$\to \pm\frac{H_{\rm st}}{\alpha}\Big[(\cos\theta_1 - 1) + \frac{1}{\overline{m}_z}\int_0^{m_{x1}} m_x\, dm_x\Big] = \pm\frac{H_{\rm st}}{\alpha}[(\cos\theta_1 - 1) + \frac{\sin^2\theta_1}{2\overline{m}_z}] \tag{22e}$$

Taking $\hat{\boldsymbol{m}}_{\rm ref} = \pm\hat{\boldsymbol{z}}$, and $1+\alpha^2 \approx 1$, Eq. (22b) employs the *z*-component of the Gilbert equation in the form of Eq. (2). Eq. (22c) employs $\boldsymbol{H}_{\rm eff}$ from Eq. (3) for the prototype example described in Eqs. (17). Eq. (22d) drops terms of order $m_y^2$, and substitutes $dt \to dm_x/(dm_x/dt)$ with $dm_x/dt$ taken from the *x*-component Gilbert equation *with both* $\boldsymbol{h}_{\rm th}$ *and* $\alpha$ *identically zero.*

This last step seems *nonphysical*, since it explicitly excludes thermal fluctuations, and manifestly ignores the fundamental *path-dependency* of the spin-torque contribution $\Delta W_{\rm st}$.

With regard to Eq. (22d), Li and Zhang further state that, "we find that the last term can be integrated out".[11] This is rigorously incorrect without specification of a path/orbit for the motion $m_y(t)$, which these authors fail to do. However, their statement is perhaps consistent within their stated approximations that $|m_y| \ll |m_x|$ and $B_s \gg H_k$, such that the 2nd term in the denominator of the integrand in Eq. (22d) can be dropped. Equivalent to letting $m_y \to 0$, this last step leads to the *meaningless* result in Eq. (22e). In either limit $H_a \to 0, \theta_1 \to \pi/2, \overline{m}_z \to 1/2$, or $H_a \to H_k, \theta_1 \to 0, \overline{m}_z \to 1$, Eq. (22e) predicts $\Delta W_{\rm st} / M_s V_{\rm fre} = 0$. This null result is not unexpected, since $\Delta W_{\rm st} \to 0$ for paths with $m_y \to 0$.[18] This author cannot further reconcile the explicit derivation of Li and Zhang[11] with that of their claimed final result.

## APPENDIX 1

For a spin-lattice system in *thermal equilibrium*, the power spectral density (PSD) matrices $\overleftrightarrow{\boldsymbol{S}}_{\boldsymbol{m}'\boldsymbol{m}'}(\omega)$ and $\overleftrightarrow{\boldsymbol{S}}_{\boldsymbol{h}'_{\rm th}\boldsymbol{h}'_{\rm th}}(\omega)$ for the thermal fluctuations of (free-layer) magnetization $\boldsymbol{m}'(t)$ or $\boldsymbol{h}'_{\rm th}(t)$ (about equilibria $\langle \boldsymbol{m}' \rangle = \langle \boldsymbol{h}'_{\rm th} \rangle = 0$) can be calculated as[12]

$$\overleftrightarrow{\boldsymbol{S}}_{\boldsymbol{m}'\boldsymbol{m}'}(\omega) = (k_B T / M_s V)[\overleftrightarrow{\chi}(\omega) - \overleftrightarrow{\chi}^{\dagger}(\omega)] / i\omega \qquad \text{(A1.1)}$$

$$\overleftrightarrow{\boldsymbol{S}}_{\boldsymbol{h}'_{\rm th}\boldsymbol{h}'_{\rm th}}(\omega) = (k_B T / M_s V)[\overleftrightarrow{\Lambda}^{\dagger}(\omega) - \overleftrightarrow{\Lambda}(\omega)] / i\omega \qquad \text{(A1.2)}$$

using the 1st or 2nd fluctuation-dissipation theorems, respectively. The relevant inverse susceptibility matrix/tensor $\overleftrightarrow{\Lambda}(\omega)$ is described in Eq. (11), with *real* $\omega$ and $H'_{xy} = H'_{yx}$. However, as noted in Sec. III, when $H_{\rm st} \propto I_b \neq 0 \Rightarrow H'_{xy} \neq H'_{yx}$, the magnetization/spin system described by Eqs.(7,11) is not strictly in thermal equilibrium with the lattice/thermal-bath, which invalidates direct application of the fluctuation-dissipation theorem (FDT). Nonetheless, if one restricts (for now) attention to $H_{\rm st} = I_b = 0$, it then follows from Eqs. (8,11,A1.1,2) that

$$\overleftrightarrow{\boldsymbol{S}}_{\boldsymbol{h}'_{\rm th}\boldsymbol{h}'_{\rm th}}(\omega) = \frac{2k_B T}{M_s V_{\rm fre}} \overleftrightarrow{\boldsymbol{D}} = \frac{2k_B T \alpha}{\gamma M_s V_{\rm fre}} \begin{pmatrix} 1 & 0 \\ 0 & 1 \end{pmatrix} \qquad \text{(A1.3)}$$

The correlation matrix $\langle \boldsymbol{h}'(t+t_0)\boldsymbol{h}(t_0)\rangle = \langle \boldsymbol{h}'(t)\boldsymbol{h}(0)\rangle \equiv (1/2\pi)\int_{-\infty}^{\infty} \overleftrightarrow{\boldsymbol{S}}_{\boldsymbol{h}'_{\rm th}\boldsymbol{h}'_{\rm th}}(\omega)e^{i\omega t}d\omega$ is defined by the inverse Fourier transform of $\overleftrightarrow{\boldsymbol{S}}_{\boldsymbol{h}'_{\rm th}\boldsymbol{h}'_{\rm th}}(\omega)$, from which Eq. (9) follows immediately.

The somewhat more complicated PSD $\overleftrightarrow{\boldsymbol{S}}_{\boldsymbol{m}'\boldsymbol{m}'}(\omega)$ requires inverting $\overleftrightarrow{\mathrm{L}}(\omega)$ to obtain $\overleftrightarrow{\mathrm{c}}(\omega)$. Taking $H'_{yx} = H'_{xy}$, the PSD $S_{m'_x m'_x}(\omega)$ for *in-plane* magnetization $m'_x$ is found to be

$$S_{m'_x m'_x}(\omega) = \frac{2k_B T\gamma\alpha}{(1+\alpha^2)M_s V_{\rm fre}} \frac{\omega^2 + \gamma^2(H'^2_{yy} + H'^2_{xy})/(1+\alpha^2)}{(\omega^2-\omega_0^2)^2 + (\omega\Delta\omega)^2}$$

$$\omega_0 \equiv \gamma\sqrt{(H'_{xx}H'_{yy} - H'_{xy}H'_{yx})/(1+\alpha^2)},\ \Delta\omega \equiv \alpha\gamma\left(H'_{yy}+H'_{xx}\right)/\left(1+\alpha^2\right) \tag{A1.4}$$

with "resonance" frequency $\omega_0$, and "line-width" $\Delta\omega$. Using typical, "ballpark" parameter values for practical systems of interest, e.g., $H'_{yy}$ ~ 10 kOe, $H'_{xx}$ ~ 0.1-1 kOe, $\alpha$ ~ 0.01-0.02, one estimates that $\Delta\omega/\omega_0$ ~ 1/10. Hence, the thermal fluctuations are, in the frequency domain, concentrated at/near the resonant frequency $\omega_0$. The time correlation function $\langle m_x(t)\,m_x(0)\rangle$ is obtainable via contour integration, and is given by

$$\langle m_x(t)\,m_x(0)\rangle = \frac{k_B T}{M_s V_{\rm fre} H'_{xx}} \exp(-|t|/\tau_c)\left[\cos(\beta'\omega_0 t) + \frac{\beta}{\beta'}\frac{H'_{yy}-H'_{xx}}{H'_{yy}+H'_{xx}}\sin(\beta'\omega_0|t|)\right]$$

$$\tau_c \equiv 2/\Delta\omega,\ \beta \equiv \tfrac{1}{2}(\Delta\omega/\omega_0),\ \beta' \equiv \sqrt{1-\beta^2} \tag{A1.5}$$

The correlations decays exponentially with characteristic "coherence" time $\tau_c$. From the prior "ballpark" parameter values, $\omega_0\tau_c$ ~ 20. Hence, the stochastic, thermally fluctuating $m_x(t)$ (or $m_y(t)$) may be roughly pictured as an oscillator undergoing quasi-periodic motion at frequency $\omega_0$ which (on average) remains phase-coherent over $\omega_0\tau_c/2\pi$ ~ 3 cycles prior to being randomly "thermalized" via spin-wave scattering and/or spin-lattice energy relaxation processes.

Finally, it was argued on physical grounds, in Sec. III, that Eq.9 (and thus Eq. A1.3) will remain valid in the presence of non-equilibrium, nonreciprocal spin-transfer torques, even if the FDT Eqs. (A1,2) are themselves are no longer valid. This assumption was also implicitly employed, perhaps without appreciation, in earlier work.[8,10]

APPENDIX 2

Within the linear approximation of Eqs. (7), the stability condition of Eq. (13) can be unified with the thermodynamic arguments of Eq. (16) as follows .Analogous to Eq. (18), the general, linearized equation of motion for $m'_y(t)$ from Eqs. (7) may be expressed as

$$(1+\alpha^2)\,dm'_y\,/\,dt+\gamma(\alpha H'_{yy}-H'_{xy})\,m'_y(t)=\gamma(H'_{xx}-\alpha H'_{yx})\,m'_x(t) \qquad \text{(A2.1)}$$

For the quasi-periodic path/orbit described in Eq. (15), at natural (resonance) frequency $\omega_0$ from Eq. (A1.4), the phase relation between $m'_{1y}$ and $m'_{1x}$ is (analogous to Eq. (19)) given by

$$\begin{gathered} m'_{1y}=[\omega_x\,/(i\omega_0+\omega_y)]\,m'_{1x} \\ \omega_0=\gamma\sqrt{H'_{xx}H'_{yy}-H'_{xy}H'_{yx}}\,,\ \omega_x=\gamma(H'_{xx}-\alpha H'_{yx})\,,\ \omega_y=\gamma(\alpha H'_{yy}-H'_{xy}) \end{gathered} \qquad \text{(A2.2)}$$

approximating $(1+\alpha^2)\cong 1$. The stability criterion of Eq. (13) may be alternatively posed physically as that condition where a (thermally excited) *natural* oscillation of the free-layer *will lose energy more quickly to spin-lattice damping then it can absorb from motion in the presence of non-zero spin-transfer torques*. The mean rate $\overline{dL/dt}$ of energy loss due to Gilbert damping is

$$\overline{dL/dt}\,/M_sV_{\text{fre}}=(\alpha/\gamma)\tfrac{\omega}{2\pi}\int_0^{2\pi/\omega}(d\boldsymbol{m}'/dt)^2\,dt=\tfrac{1}{2}(\alpha/\gamma)\,\omega_0^2\,(m'^*_{1x}m'_{1x}+m'^*_{1y}m'_{1y}) \qquad \text{(A2.3)}$$

The energy loss-rate stability criterion is $\overline{dL/dt}>-\overline{dW/dt}$, where $-\overline{dW/dt}$ from Eq. (16) is the mean rate of energy *absorption by* the free-layer *from* the *dc* spin-polarized current. Substituting Eqs. (A2.2) into both Eq. (16) and Eq. (A2.3), one then obtains the stability criterion

$$\begin{gathered} \overline{dL/dt}+\overline{dW/dt}>0 \\ \to(\alpha/\gamma)\,\omega_0^2\,(\omega_0^2+\omega_x^2+\omega_y^2)/(\omega_0^2+\omega_y^2)+(H'_{yx}-H'_{xy})\,\omega_0^2\omega_x\,/(\omega_0^2+\omega_y^2)\ >0 \\ \to(\alpha/\gamma)(\omega_0^2+\omega_x^2+\omega_y^2)+(H'_{yx}-H'_{xy})\,\omega_x\ >0 \end{gathered} \qquad \text{(A2.4)}$$

Keeping terms to *first* order in $\alpha$, and substituting from Eqs. (A2.2), the last result becomes

$$\alpha(H'_{xx}H'_{yy}+H'^2_{xx}+H'^2_{xy}-H'^2_{yx})+H'_{xx}(H'_{yx}-H'_{xy})\ >0 \qquad \text{(A2.5)}$$

Invoking (without loss of generality) the choice of "principle-axes" so that $H'_{xy}=-H'_{yx}\propto H_{\text{st}}$, and physically requiring $H'_{xx}>0$ by Eq. (12), the result of Eq. (A2.5) immediately becomes

$$\alpha(H'_{yy}+H'_{xx})+(H'_{yx}-H'_{xy})\ >0 \qquad \text{(A2.6)}$$

which is identical to Eq. (13), and hence supports the interpretation of the work $\Delta W$ in Sec. IV.

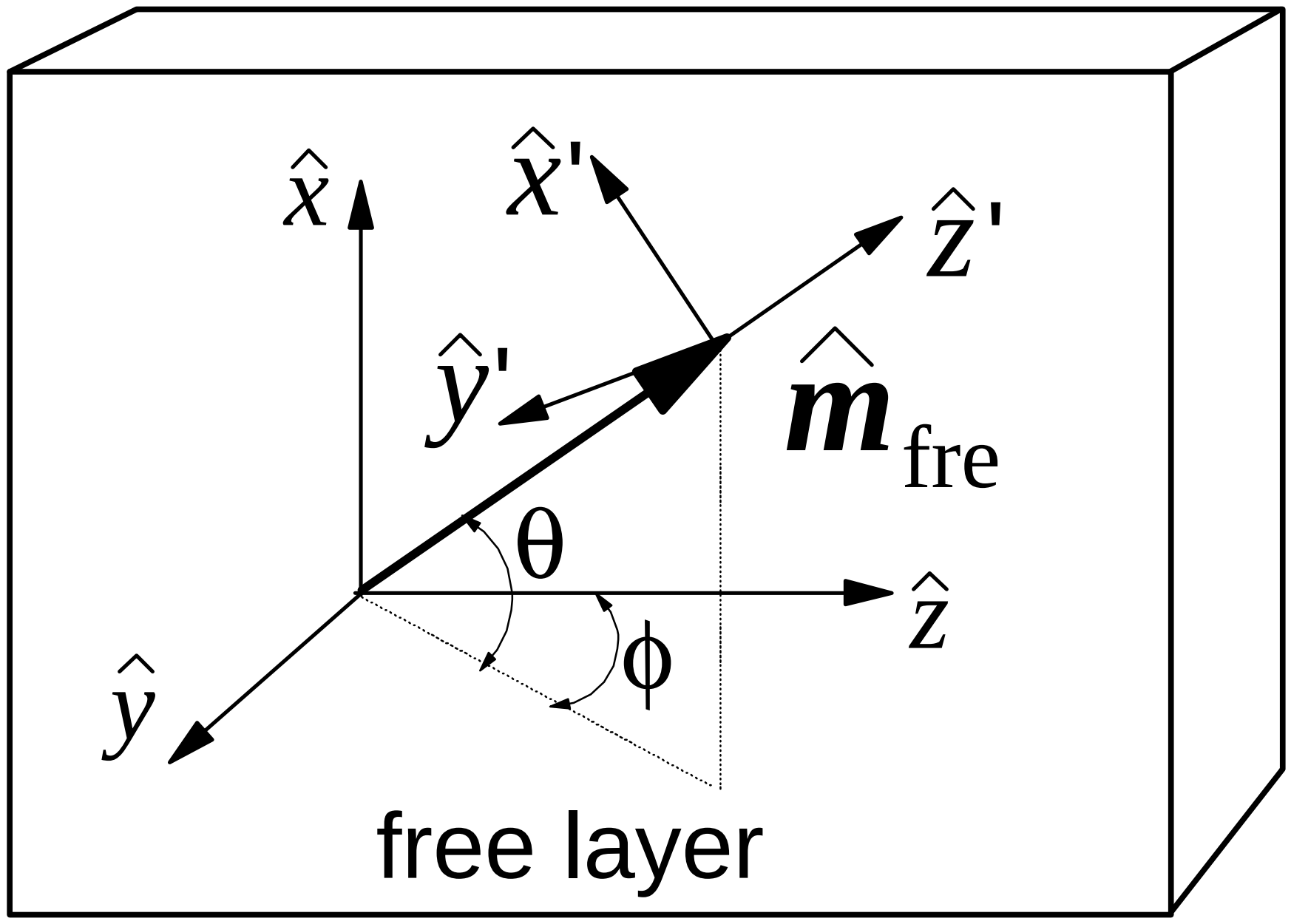

Fig. 1 Simple cartoon of the thin-film free-layer (film-plane = *x*-*z* plane), defining/illustrating the rotational transformation described in Eq. (5). Throughout this article, the free-layer is treated for simplicity as a single-domain magnetic "particle", or macro-spin of orientation $\hat{\boldsymbol{m}}_{\text{fre}} \Leftrightarrow \hat{\boldsymbol{m}}$ .